\documentclass
[amsmath,amssymb,superscriptaddress,pra,showpacs]{revtex4}

\usepackage{epsfig}
\usepackage{amsmath}
\usepackage{graphicx}
\usepackage{bm}
\usepackage{ulem}

\begin{document}

\title{Radiation-pressure self-cooling of a micromirror in a cryogenic environment\\}

\author{Simon Gr\"oblacher}
\affiliation{Institute for Quantum Optics and Quantum Information (IQOQI), Austrian Academy of Sciences, Boltzmanngasse 3, A--1090 Vienna, Austria}

\author{Sylvain Gigan}
\altaffiliation[Permanent address: ]{Laboratoire Photon et Mati\`{e}re, Ecole Superieure de Physique et de Chimie Industrielles, CNRS-UPR A0005, 10 rue Vauquelin, 75005 Paris, France}
\affiliation{Institute for Quantum Optics and Quantum Information (IQOQI), Austrian Academy of Sciences, Boltzmanngasse 3, A--1090 Vienna, Austria}

\author{Hannes R. B\"ohm}
\affiliation{Institute for Quantum Optics and Quantum Information (IQOQI), Austrian Academy of Sciences, Boltzmanngasse 3, A--1090 Vienna, Austria}
\affiliation{Faculty of Physics, University of Vienna, Boltzmanngasse 5, A--1090 Vienna, Austria}

\author{Anton Zeilinger}
\affiliation{Institute for Quantum Optics and Quantum Information (IQOQI), Austrian Academy of Sciences, Boltzmanngasse 3, A--1090 Vienna, Austria}
\affiliation{Faculty of Physics, University of Vienna, Boltzmanngasse 5, A--1090 Vienna, Austria}

\author{Markus Aspelmeyer}
\affiliation{Institute for Quantum Optics and Quantum Information (IQOQI), Austrian Academy of Sciences, Boltzmanngasse 3, A--1090 Vienna, Austria}

\begin{abstract}
We demonstrate radiation-pressure cavity-cooling of a mechanical mode of a micromirror starting from cryogenic temperatures. To achieve that, a high-finesse Fabry-P\'{e}rot cavity ($F\approx 2200$) was actively stabilized inside a continuous-flow $^{4}$He cryostat. We observed optical cooling of the fundamental mode of a $50~\mu$m$\times50~\mu$m$\times5.4~\mu$m singly-clamped micromirror at $\omega_m=3.5$~MHz from 35~K to approx.\ 290~mK. This corresponds to a thermal occupation factor of $\langle n\rangle\approx 1\times10^4$. The cooling performance is only limited by the mechanical quality and by the optical finesse of the system. Heating effects, \textit{e.g.} due to absorption of photons in the micromirror, could not be observed. These results represent a next step towards cavity-cooling a mechanical oscillator into its quantum ground state~\cite{Groeblacher2008a}.
\end{abstract}


\maketitle

Optomechanical interactions in high-finesse cavities offer a new promising route for the ongoing experimental efforts to achieve the quantum regime of massive mechanical systems~\cite{LaHaye2004,Schwab2005}. They allow to cool mechanical degrees of freedom of movable mirrors via radiation-pressure backaction~\cite{Braginsky2002}, in principle even into their quantum ground state~\cite{Marquardt2007,Wilson-Rae2007,Genes2007}. The working principle of this cooling method has been demonstrated in a series of recent experiments~\cite{Gigan2006,Arcizet2006b,Schliesser2006,Corbitt2007}. Ground-state cooling will eventually require to realize the scheme in a cryogenic environment. Optomechanical feedback cooling~\cite{Mancini1998,Cohadon1999,Kleckner2006b,Arcizet2006a,Poggio2007}, another quantum limited strategy~\cite{Courty2001,Vitali2002,Genes2007}, has recently taken this step by demonstrating cooling of a 3.8~kHz resonator mode from a starting temperature of 2~K to an effective noise temperature of 2.9~mK (or $\langle n\rangle\approx 2.1\times10^4$)~\cite{Poggio2007}. To achieve and surpass such a performance for radiation-pressure backaction schemes requires stable operation of a high-finesse cavity inside a cryostat~\cite{Tittonen1999} and sufficiently strong optomechanical coupling~\cite{Gigan2006,Arcizet2006b,Schliesser2006,Corbitt2007}. Here we report the combination of these requirements in a single experiment using a high-reflectivity micromechanical resonator. We observe radiation-pressure backaction cooling of the fundamental mode of the micromirror at $\omega_m/2\pi=557$~kHz from 35~K to 290~mK (or $\langle n\rangle\approx 1\times10^4$), limited only by the optical finesse of the cavity and by the mechanical quality of the micromirror.\\

How does radiation-pressure cooling work? The basic setup comprises an optical cavity of frequency $\omega_c$, pumped by a laser at frequency $\omega_l$, that is bounded by a mechanical oscillator of resonance frequency $\omega_m$. By reflecting photons off the mechanical resonator, in our case a movable micromirror, the intracavity field exerts a radiation-pressure force on the mechanical system. Detuning of the optical cavity ($\Delta=\omega_c-\omega_l\neq 0$) can result in a net positive~($\Delta<0$) or negative~($\Delta>0$) energy transfer from the radiation field to the mechanical oscillator, corresponding to either heating or cooling of the mechanical mode. There are different views to understand the cooling effect. Considering the full dynamics of the system, radiation-pressure forces in a detuned cavity behave as a viscous force that modifies the mechanical susceptibility~\cite{Braginsky2002,Metzger2004,Arcizet2006c}. Cooling occurs as a consequence of the delayed (retarded) force response to thermal fluctuations of the mechanical resonator, which is caused by the finite cavity decay rate $\kappa$. It is worth noting that retardation-based optomechanical cooling is not restricted to radiation pressure and its principle was in fact for the first time demonstrated using photothermal forces~\cite{Metzger2004}. Going beyond (semi-)classical descriptions, a full quantum treatment~\cite{Paternostro2006,Wilson-Rae2007,Vitali2007,Marquardt2007} can provide an interesting interpretation of the cooling effect as quantum state transfer between two oscillators, \textit{i.e.} the cavity field and the mechanical mode~\cite{Zhang2003}. This is related to the thermodynamic analogy, by which an entropy flow occurs from the thermally excited mechanical mode to the low-entropy laser field. Finally, the comparison of the photon-phonon interaction with three-wave mixing leads to the intuitive picture of sideband-cooling~\cite{Marquardt2007,Wilson-Rae2007}, as is well known from laser-cooling of atoms and ions~\footnote{Note that in our case radiation pressure originates from the reflection of photons off the mirror surface and not from absorption and re-emission as is the case in conventional laser cooling. Still, the cooling mechanism of both schemes is completely analogous.}.\\

\begin{figure}[htbp]
\centerline{\includegraphics[width=0.8\textwidth]{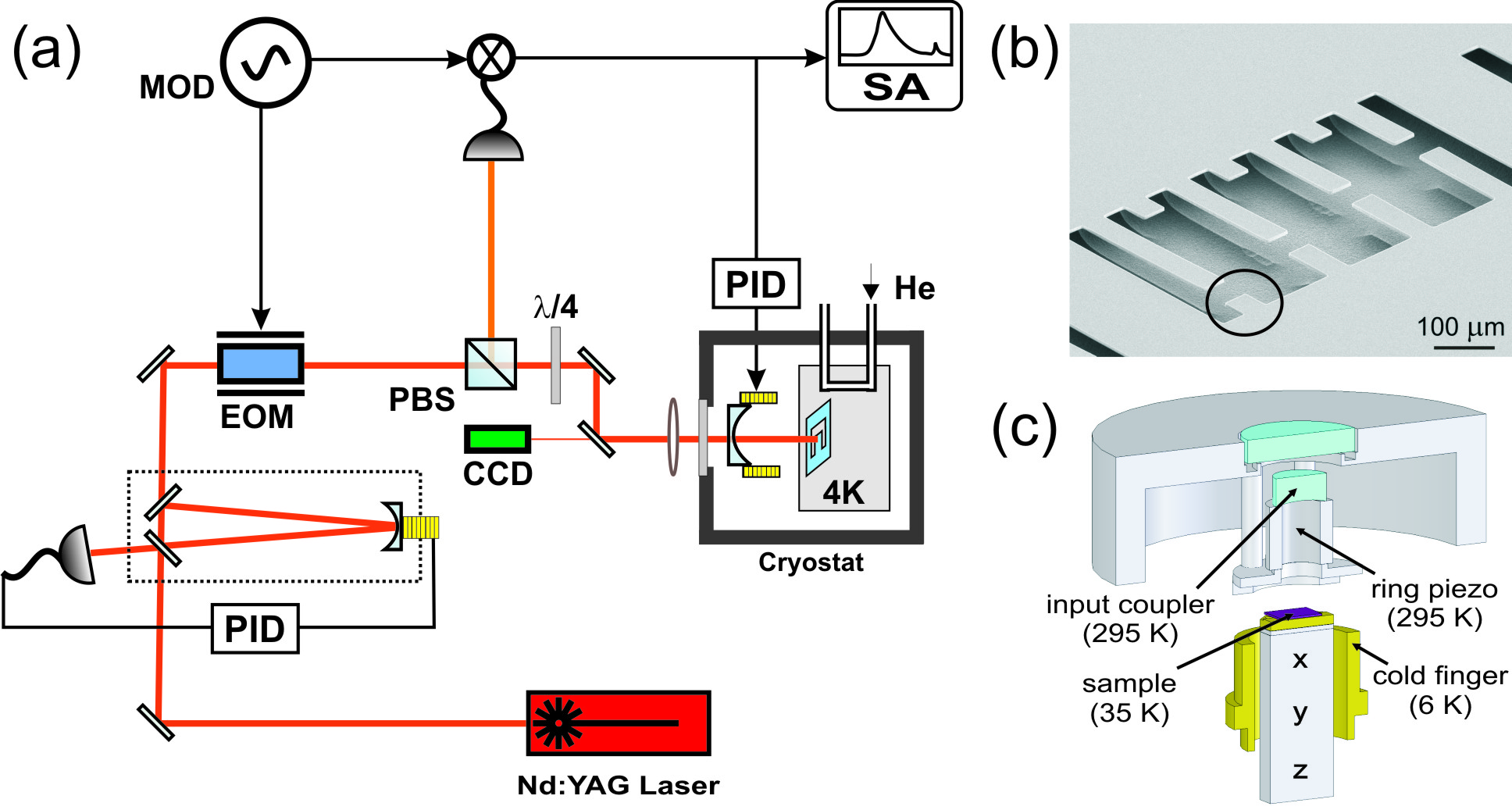}} \caption{Experimental Scheme.\ (a) The pump beam is spatially and spectrally filtered in a ring cavity locked to the laser frequency. After phase modulation using an electro-optic modulator (EOM), for Pound-Drever-Hall (PDH) locking, the pump is injected into the micromirror Fabry-P\'{e}rot (FP) cavity, which is mounted inside a $^4$He cryostat. The beam reflected from the FP cavity is detected behind a polarizing beam splitter (PBS). The PDH signal is obtained by demodulating the detected signal by the EOM driving frequency and is used for actively stabilizing the cavity length and for monitoring the dynamics of the mechanical mode. Alignment is done via a CCD camera.\ (b) SEM picture of a group of micromirrors.\ (c) Cavity mounting inside the cryostat (see text).} \label{fig1}
\end{figure}

Our mechanical objects are oscillating micromirrors of high reflectivity that consist solely of a dielectric Bragg-mirror coating~\cite{Boehm2006}. Compared to our previous work~\cite{Gigan2006} we have used a different coating material to achieve both higher reflectivity and lower inherent absorption. This allowed us to increase the radiation-pressure coupling and to avoid residual photothermal effects. For the fabrication process we start from a high-reflectivity coating ($R > 0.9999$) made out of 40 alternating layers of $Ta_2O_5$ and $SiO_2$ deposited on silicon. We used reactive ion etching to define the resonator shape and selective dry etching of the substrate to free the structures. All mechanical resonators form singly clamped cantilevers with a thickness of $5.4~\mu$m, a width of $50~\mu$m and a length between $50~\mu$m and $300~\mu$m (fig.~\ref{fig1}b). We found mechanical quality factors $Q\approx 1000 - 3000$ and reflectivities of $R > 0.9999$.\\

The full experimental setup is sketched in fig.~\ref{fig1}a. We use the micromirror as an end mirror in a high-finesse Fabry-P\'{e}rot (FP) cavity, which is pumped by a an ultrastable Nd:YAG laser operating in continuous-wave mode at a wavelength of 1064~nm. The input coupler of the FP cavity is a concave massive mirror (radius of curvature:\ $25$~mm; reflectivity at 1064 nm:\ $0.9993$) that is attached to a ring piezo (PZT) in order to actively modify the cavity length. We chose the length $L$ of the cavity slightly shorter than for the semi-concentric case ($L=25$~mm) in order to have a stable cavity and a small cavity-mode waist $w_0$ on the micromirror ($w_0\approx 10~\mu$m). The cavity is mounted inside a continuous-flow $^{4}$He cryostat (fig.~\ref{fig1}c). The input coupler is attached to the outer shield of the cryostat and therefore always maintains at room temperature. The silicon wafer that holds the micromirrors is glued on a sample holder that is in thermal contact with the cryostat cold finger. A 3-axis translation stage allows precise positioning of the micromirror on the chip with respect to the footprint of the cavity beam. We monitor both position and size of the cavity mode via an external imaging system. In operation, the cryostat is first evacuated to $10^{-6}$~mbar. Cryogenic cooling is achieved by a continuous flow of helium in direct contact with the cold finger. The additional cryogenic freeze out reduces the pressure to below $3\times 10^{-7}$~mbar. On cooling the cryostat from room temperature to approx.\ 6~K (measured temperature at the cold finger), the thermal contraction of the cavity ($1-2$~mm in total) can be compensated by the 3-axis translation stage. The temperature of the sample holder is monitored via an additional sensor directly attached to it. For a measured cold-finger temperature of 6~K we observe a sample holder temperature of approx.\ 20~K and an actual sample temperature of 35~K, which we infer from the calibrated power spectrum of the micromirror motion as mode temperature at zero optical detuning (see below). We attribute the temperature gradient to heating of the sample by blackbody radiation from the input coupler, which is kept at 295~K only a few millimeters away from the sample, in combination with finite thermal conductivity between sample, sample holder and cold finger. Both at room temperature and at cryogenic temperatures we observe stable locking of the cavity for a finesse of up to $8000$. We achieve typical mode matching efficiencies into the cavity of $80\%$. \\

\begin{figure}[htbp]
\centerline{\includegraphics[width=0.6\textwidth]{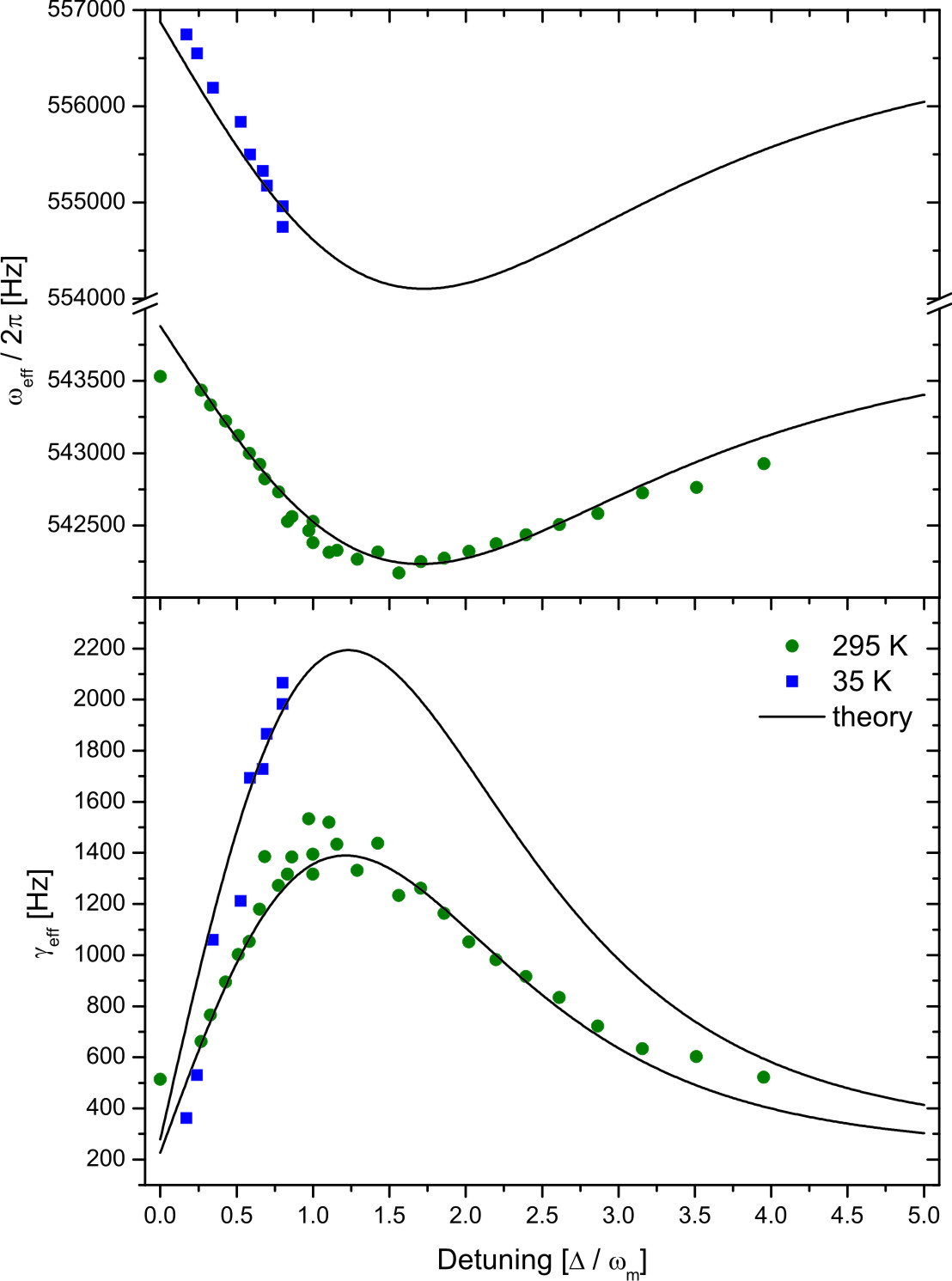}} \caption{Modified micromirror dynamics due to cavity detuning. Shown is the micromirror's effective frequency $\omega_{eff}/2\pi$ and effective damping $\gamma_{eff}$ both at room temperature and at 35~K for various detuning values at a laser power of 1~mW. Maximal cooling is obtained approximately at a detuning of $\omega_{m}$, where the net phonon transfer to the optical field is maximized. The solid lines are fits to the data based on the semi-classical model for radiation-pressure backaction (see text).} \label{fig2}
\end{figure}

\begin{figure}[htbp]
\centerline{\includegraphics[width=0.6\textwidth]{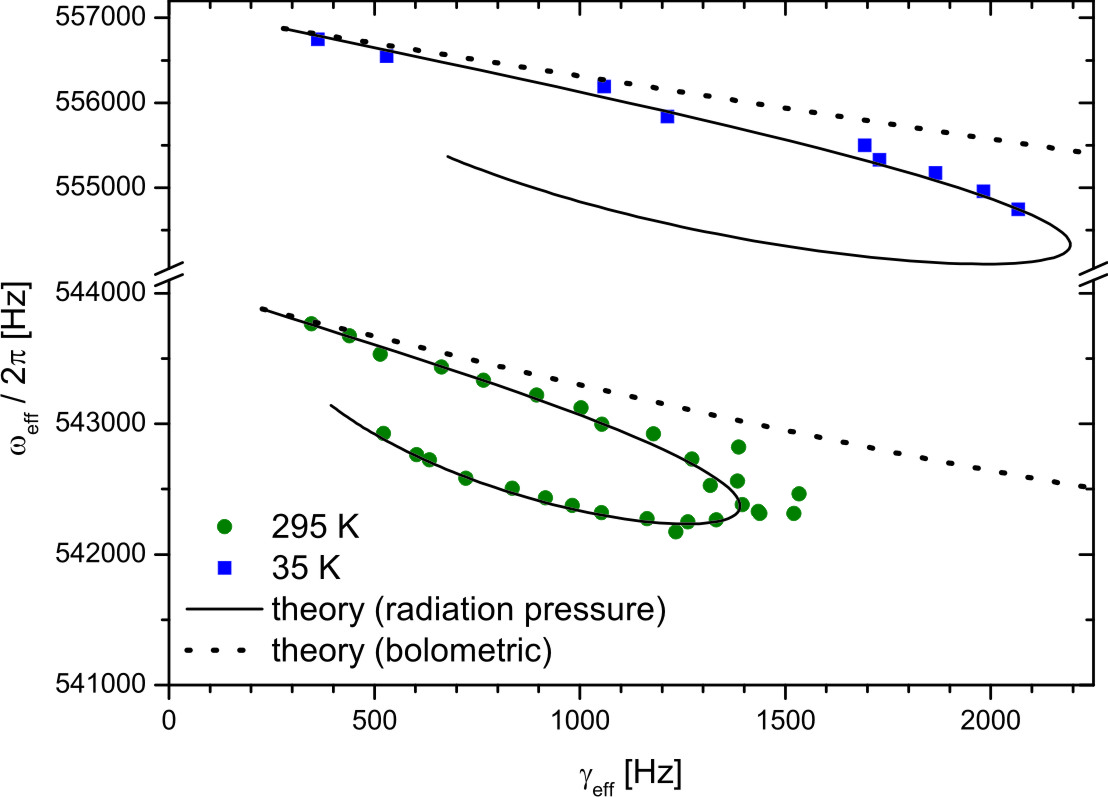}} \caption{Radiation-pressure backaction. The data follow the curve from the top-left to the bottom-left. The solid lines are fits to the data based on the semi-classical model for pure radiation-pressure backaction (see text). The dotted lines show the expected behavior for bolometric (photothermal) forces when using the same parameters. Even at low temperature a clear deviation from photo-thermal behavior is observed and the data is well described by radiation-pressure effects.} \label{fig3}
\end{figure}

To observe the desired backaction cooling we monitor the dynamics of the different eigenmodes of the micromirror vibration by measuring its displacement power spectrum $S_x(\omega)$~\cite{Paternostro2006}. This is done by analyzing the Pound-Drever-Hall (PDH) signal in the light backreflected from the FP cavity~\cite{Tittonen1999,Arcizet2006a,Gigan2006}, a method which is based on the interference of phase-modulated side bands of the pump laser~\cite{Gigan2006,Arcizet2006b}. The main idea is that the PDH error signal of a locked cavity is proportional to the cavity length. While we use the low-frequency part of the PDH signal as an error signal to actively stabilize the cavity length to the wanted detuning $\Delta$, the high frequency part is directly proportional to the displacement power spectrum $S_x$ of the micromirror~\cite{Paternostro2006}~\footnote{The ratio between PDH power spectrum and displacement power spectrum $S_x$ depends on the cavity detuning $\Delta$. We can eliminate the unwanted detuning dependence by normalizing $S_x$ via a reference signal of a known constant displacement power spectrum $S_{ref}$ that is generated by frequency modulation of the pump laser. In addition, $S_{ref}$ is an absolute calibration of the effective mass of the mechanical oscillator, as is outlined in detail \textit{e.g.} in ~\cite{Gigan2006}.}. One can evaluate the effective mode temperature via the area of the measured power spectrum as $T_{eff}=\frac{m\omega_0^2}{k_B}\langle x^2\rangle$ ($m$:\ effective mass at the probing point, $\omega_0$:\ mode frequency, $k_B$:\ Boltzmann's constant, $\langle x^2\rangle=\int_{-\infty}^{+\infty}d\omega S_x(\omega)$). \\
Backaction cooling is accompanied by a modified dynamics of the mechanical mode, specifically by a shift both in resonance frequency $\omega_{eff}$ and in damping $\gamma_{eff}$. This can be used to identify the nature of the backaction force:\ for a known effective mass and optical pump power, radiation-pressure forces are uniquely determined by the time dependence of the cavity decay and can therefore be distinguished from forces of dissipative nature such as photothermal forces~\cite{Marquardt2007}. We obtain these effective values directly via the power spectrum $S_x$, which, for a classical harmonic oscillator, is given by
\begin{equation}
S_x(\omega)=\frac{4 k_B T \gamma_0}{\pi m}\frac{1}{(\omega_{eff}^2-\omega^2)^2+4 \gamma_{eff}^2\omega^2},
\label{eq:Sx}
\end{equation}
where $\gamma_0$ is the mechanical damping of the unperturbed mechanical oscillator, \textit{i.e.} the damping at zero detuning. To minimize radiation pressure effects we used very low input power ($\approx 30~\mu$W) and probed the mode at a point of high effective mass, \textit{i.e.} close to a node of vibration. The values for $\gamma_{eff}$ and $\omega_{eff}$ were obtained from fits to the measured power spectra using eq.~(\ref{eq:Sx}).\\

We first confirmed that our optomechanical system is dominated by radiation-pressure backaction. For that purpose, we monitor the modified dynamics of the mechanical mode of a micromirror and compare it with the theoretical predictions for radiation-pressure effects. The results for various cavity detunings are shown in fig.~\ref{fig2}. The solid lines are fits to the data using the semi-classical approach described in~\cite{Arcizet2006c}. We obtain a fitted cavity finesse $F=2300$ and a fitted effective mass of $m=125$~ng. These values are consistent with our independent estimate of $F=2800\pm600$ and $m=(110\pm30)$~ng (obtained from $S_{ref}$)~\footnote{The reduction in finesse compared to the value of $8000$ is due to our choice of the optimal working point on the cantilever close to the tip of the micromirror, where edge diffraction increased the losses in the cavity.}. Note that the finesse is measured by slowly scanning the cavity length. The corresponding measurement uncertainty arises from mechanical vibrations of the setup. We also performed a measurement on the mirror at 35~K (fig.~\ref{fig2}), however with a reduced detuning range (for technical reasons the full detuning range was not available at low temperature). Again, the fit values of $F=2200$ and $m=40$~ng are consistent with our estimates of $F=2800\pm800$ and $m=(30\pm10)$~ng and therefore confirm the radiation-pressure nature of the interaction. In contrast to radiation-pressure forces, photothermal forces are always subject to an exponential retardation due to the dissipative nature of the force and therefore produce a different dynamics on detuning~\cite{Marquardt2007}. We have used the same parameters to simulate the expected behavior resulting from such a force (fig.~\ref{fig3}), which can clearly not serve as an explanation for our data.\\

\begin{figure}[htbp]
\centerline{\includegraphics[width=0.6\textwidth]{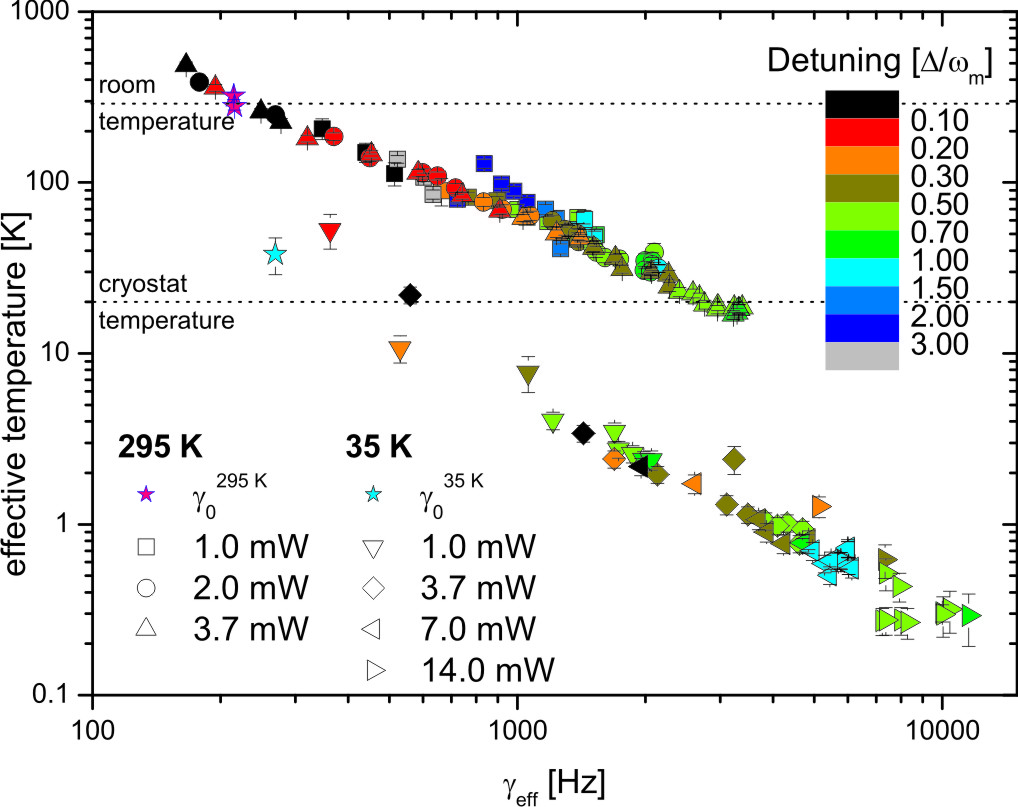}} \caption{Radiation-pressure cooling in a cryogenic high-finesse cavity. Shown are effective temperature $T_{eff}$ and the effective damping $\gamma_{eff}$ in a detuned cavity for various laser powers. Different laser powers correspond to different symbols. Values of detuning (in units of $\omega_m$) are encoded in color. Starting from cryogenic temperatures (the given cryostat temperature is the measured sample holder temperature) we observe backaction cooling down to 290~mK (or $\langle n\rangle\approx1\times10^4$). The cooling performance is not limited by heating but by optical finesse and mechanical quality factor of the optomechanical system.} \label{fig4}
\end{figure}

Finally, we demonstrate radiation-pressure backaction cooling in a cryogenic cavity. Figure~\ref{fig4} shows measurements performed on the fundamental mechanical mode at $\omega_m=2\pi\times 557$~kHz of the micromirror. For each detuning and optical power level we obtained $\langle x^2\rangle$, $\omega_{eff}$ and $\gamma_{eff}$ directly from the fits to the measured displacement spectrum $S_x$. The effective mass $m_{eff}$ is obtained as described in the previous paragraph by fitting the data sets of same optical power (at a given cryostat temperature) using a semi-classical approach to radiation-pressure backaction.
The effective temperature is obtained by plotting $m_{eff}\cdot\omega_{eff}^2\cdot\langle x^2\rangle$ normalized to the value obtained at zero detuning at room temperature (295~K). When cooling the cavity down to a sample holder temperature of 20~K we find a measured mode temperature at zero detuning (corresponding to $\gamma_0^{35K}=2\pi\times 269$~Hz) of approximately 35~K. On detuning, the mode temperature decreases as expected for both starting temperatures. For a given laser power the effective mode temperature decreases with increasing detuning until $\Delta\approx \omega_m$, where the cooling is optimal. The effective temperature increases again on further increasing the detuning. When starting from room temperature we observe a minimum temperature of approximately 17~K at an input laser power of 3.7~mW. Starting with a cryogenic cavity we observe a minimum mode temperature of approximately 290~mK for 14~mW laser power. This corresponds to a thermal occupation factor of $\langle n\rangle\approx1\times10^4$.\\

The cooling performance is not limited by residual heating effects. In the ideal (semi-)classical case $T_{eff}\approx T_{0}\frac{\gamma_0}{\gamma_{eff}}$ (for $\omega_{eff}\ll\gamma_{eff}$ and $T_0$:\ environment temperature), as one can see from integrating eq.~(\ref{eq:Sx}) and by using the equipartition theorem. We observe this behavior as linear dependence on the double-logarithmic scale of fig.~\ref{fig4}. In case of heating, \textit{e.g.} by absorption of photons, one would expect a dependence of the mode temperature on the laser power even for the same effective damping $\gamma_{eff}$. In other words, data points taken at different laser powers would not fall on the same line. The fact that we observe no deviation from the linear dependence for increasing laser power indicates that no significant heating of the mode occurs. We should also note that our experimental parameters ($F=2200, \omega_m=3.5\times10^6$) fulfill the threshold condition for ground-state cooling, because $\omega_{m}/\kappa=0.2>1/\sqrt{32}$~\cite{Wilson-Rae2007}. Our present cooling performance is only limited by the initial temperature $T_0$ of the environment, \textit{i.e.} the performance of the cryostat, and by the achieved damping ratio $\frac{\gamma_0}{\gamma_{eff}}$. Future improvements will have to include a further reduction of $T_0$, \textit{e.g.} by including a radiation shield to protect the sample from blackbody radiation, a decrease in $\gamma_0$, \textit{i.e.} a larger mechanical Q, and an increase of optical intracavity power, in particular via an increase of finesse.

We have demonstrated radiation-pressure backaction cooling of a micromirror in a high-finesse cavity at cryogenic temperatures. Starting from a sample temperature of approximately 35~K we achieve an effective mode temperature of 290~mK ($\langle n\rangle\approx 1\times 10^4$), limited only by the micromirror's mechanical quality factor and by its optical reflectivity. We consider this a next step towards exploiting the rich structure promised by optomechanical systems when entering the mechanical quantum regime~\cite{Bose1997,Pinard2005b,Pirandola2006,Vitali2007}. We believe that the combination of cryogenic cooling with (active or passive) feedback techniques~\cite{Naik2006,Poggio2007,Thompson2007} will be an essential step to achieve this goal.

\acknowledgments
We are grateful to J. B. Hertzberg and K. Schwab for valuable support in sample preparation, and to K. Gugler, T. Paterek, M. Paternostro and D. Vitali for discussion. We acknowledge financial support by the FWF (Projects P19539-N20 and L426-N20), by the IST funded Integrated Project QAP (Contract 015846) of the European Commission, by the City of Vienna and by the Foundational Questions Institute fqxi.org (Grant RFP1-06-14). S. Gr\"oblacher is recipient of a DOC-fellowship of the Austrian Academy of Sciences.


\begin{thebibliography}{29}
\expandafter\ifx\csname natexlab\endcsname\relax\def\natexlab#1{#1}\fi
\expandafter\ifx\csname bibnamefont\endcsname\relax
  \def\bibnamefont#1{#1}\fi
\expandafter\ifx\csname bibfnamefont\endcsname\relax
  \def\bibfnamefont#1{#1}\fi
\expandafter\ifx\csname citenamefont\endcsname\relax
  \def\citenamefont#1{#1}\fi
\expandafter\ifx\csname url\endcsname\relax
  \def\url#1{\texttt{#1}}\fi
\expandafter\ifx\csname urlprefix\endcsname\relax\def\urlprefix{URL }\fi
\providecommand{\bibinfo}[2]{#2}
\providecommand{\eprint}[2][]{\url{#2}}

\bibitem[{\citenamefont{in}(2008)\citenamefont{published}}]{Groeblacher2008a}
\bibinfo{author}{\bibfnamefont{This work was published}~\bibnamefont{in}}
  \bibinfo{journal}{Europhys.\ Lett.} \textbf{\bibinfo{volume}{81}},
  \bibinfo{pages}{54003} (\bibinfo{year}{2008}).

\bibitem[{\citenamefont{LaHaye et~al.}(2004)\citenamefont{LaHaye, Buu,
  Camarota, and Schwab}}]{LaHaye2004}
\bibinfo{author}{\bibfnamefont{M.~D.} \bibnamefont{LaHaye}},
  \bibinfo{author}{\bibfnamefont{O.}~\bibnamefont{Buu}},
  \bibinfo{author}{\bibfnamefont{B.}~\bibnamefont{Camarota}}, \bibnamefont{and}
  \bibinfo{author}{\bibfnamefont{K.~C.} \bibnamefont{Schwab}},
  \bibinfo{journal}{Science} \textbf{\bibinfo{volume}{304}},
  \bibinfo{pages}{74} (\bibinfo{year}{2004}).

\bibitem[{\citenamefont{Schwab and Roukes}(2005)}]{Schwab2005}
\bibinfo{author}{\bibfnamefont{K.~C.} \bibnamefont{Schwab}} \bibnamefont{and}
  \bibinfo{author}{\bibfnamefont{M.~L.} \bibnamefont{Roukes}},
  \bibinfo{journal}{Physics Today}  (\bibinfo{year}{2005}).

\bibitem[{\citenamefont{Braginsky and Vyatchanin}(2002)}]{Braginsky2002}
\bibinfo{author}{\bibfnamefont{V.}~\bibnamefont{Braginsky}} \bibnamefont{and}
  \bibinfo{author}{\bibfnamefont{S.~P.} \bibnamefont{Vyatchanin}},
  \bibinfo{journal}{Phys.\ Lett.\ A} \textbf{\bibinfo{volume}{293}},
  \bibinfo{pages}{228} (\bibinfo{year}{2002}).

\bibitem[{\citenamefont{Marquardt et~al.}(2007)\citenamefont{Marquardt, Chen,
  Clerk, and Girvin}}]{Marquardt2007}
\bibinfo{author}{\bibfnamefont{F.}~\bibnamefont{Marquardt}},
  \bibinfo{author}{\bibfnamefont{J.~P.} \bibnamefont{Chen}},
  \bibinfo{author}{\bibfnamefont{A.~A.} \bibnamefont{Clerk}}, \bibnamefont{and}
  \bibinfo{author}{\bibfnamefont{S.~M.} \bibnamefont{Girvin}},
  \bibinfo{journal}{Phys.\ Rev.\ Lett.} \textbf{\bibinfo{volume}{99}},
  \bibinfo{pages}{093902} (\bibinfo{year}{2007}).

\bibitem[{\citenamefont{Wilson-Rae et~al.}(2007)\citenamefont{Wilson-Rae,
  Nooshi, Zwerger, and Kippenberg}}]{Wilson-Rae2007}
\bibinfo{author}{\bibfnamefont{I.}~\bibnamefont{Wilson-Rae}},
  \bibinfo{author}{\bibfnamefont{N.}~\bibnamefont{Nooshi}},
  \bibinfo{author}{\bibfnamefont{W.}~\bibnamefont{Zwerger}}, \bibnamefont{and}
  \bibinfo{author}{\bibfnamefont{T.~J.} \bibnamefont{Kippenberg}},
  \bibinfo{journal}{Phys.\ Rev.\ Lett.} \textbf{\bibinfo{volume}{99}},
  \bibinfo{pages}{093901} (\bibinfo{year}{2007}).

\bibitem[{\citenamefont{Genes et~al.}(2007)\citenamefont{Genes, Vitali,
  Tombesi, Gigan, and Aspelmeyer}}]{Genes2007}
\bibinfo{author}{\bibfnamefont{C.}~\bibnamefont{Genes}},
  \bibinfo{author}{\bibfnamefont{D.}~\bibnamefont{Vitali}},
  \bibinfo{author}{\bibfnamefont{P.}~\bibnamefont{Tombesi}},
  \bibinfo{author}{\bibfnamefont{S.}~\bibnamefont{Gigan}}, \bibnamefont{and}
  \bibinfo{author}{\bibfnamefont{M.}~\bibnamefont{Aspelmeyer}},
  \bibinfo{journal}{arXiv:0705.1728}  (\bibinfo{year}{2007}).

\bibitem[{\citenamefont{Gigan et~al.}(2006)\citenamefont{Gigan, B\"ohm,
  Paternostro, Blaser, Langer, Hertzberg, Schwab, B\"auerle, Aspelmeyer, and
  Zeilinger}}]{Gigan2006}
\bibinfo{author}{\bibfnamefont{S.}~\bibnamefont{Gigan}},
  \bibinfo{author}{\bibfnamefont{H.~R.} \bibnamefont{B\"ohm}},
  \bibinfo{author}{\bibfnamefont{M.}~\bibnamefont{Paternostro}},
  \bibinfo{author}{\bibfnamefont{F.}~\bibnamefont{Blaser}},
  \bibinfo{author}{\bibfnamefont{G.}~\bibnamefont{Langer}},
  \bibinfo{author}{\bibfnamefont{J.~B.} \bibnamefont{Hertzberg}},
  \bibinfo{author}{\bibfnamefont{K.~C.} \bibnamefont{Schwab}},
  \bibinfo{author}{\bibfnamefont{D.}~\bibnamefont{B\"auerle}},
  \bibinfo{author}{\bibfnamefont{M.}~\bibnamefont{Aspelmeyer}},
  \bibnamefont{and}
  \bibinfo{author}{\bibfnamefont{A.}~\bibnamefont{Zeilinger}},
  \bibinfo{journal}{Nature} \textbf{\bibinfo{volume}{444}}, \bibinfo{pages}{67}
  (\bibinfo{year}{2006}).

\bibitem[{\citenamefont{Arcizet
  et~al.}(2006{\natexlab{a}})\citenamefont{Arcizet, Cohadon, Briant, Pinard,
  and Heidmann}}]{Arcizet2006b}
\bibinfo{author}{\bibfnamefont{O.}~\bibnamefont{Arcizet}},
  \bibinfo{author}{\bibfnamefont{P.~F.} \bibnamefont{Cohadon}},
  \bibinfo{author}{\bibfnamefont{T.}~\bibnamefont{Briant}},
  \bibinfo{author}{\bibfnamefont{M.}~\bibnamefont{Pinard}}, \bibnamefont{and}
  \bibinfo{author}{\bibfnamefont{A.}~\bibnamefont{Heidmann}},
  \bibinfo{journal}{Nature} \textbf{\bibinfo{volume}{444}}, \bibinfo{pages}{71}
  (\bibinfo{year}{2006}{\natexlab{a}}).

\bibitem[{\citenamefont{Schliesser et~al.}(2006)\citenamefont{Schliesser,
  Del'Haye, Nooshi, Vahala, and Kippenberg}}]{Schliesser2006}
\bibinfo{author}{\bibfnamefont{A.}~\bibnamefont{Schliesser}},
  \bibinfo{author}{\bibfnamefont{P.}~\bibnamefont{Del'Haye}},
  \bibinfo{author}{\bibfnamefont{N.}~\bibnamefont{Nooshi}},
  \bibinfo{author}{\bibfnamefont{K.~J.} \bibnamefont{Vahala}},
  \bibnamefont{and} \bibinfo{author}{\bibfnamefont{T.~J.}
  \bibnamefont{Kippenberg}}, \bibinfo{journal}{Phys.\ Rev.\ Lett.}
  \textbf{\bibinfo{volume}{97}}, \bibinfo{pages}{243905}
  (\bibinfo{year}{2006}).

\bibitem[{\citenamefont{Corbitt et~al.}(2007)\citenamefont{Corbitt, Chen,
  Innerhofer, M\"uller-Ebhardt, Ottaway, Rehbein, Sigg, Whitcomb, Wipf,
  et~al.}}]{Corbitt2007}
\bibinfo{author}{\bibfnamefont{T.}~\bibnamefont{Corbitt}},
  \bibinfo{author}{\bibfnamefont{Y.}~\bibnamefont{Chen}},
  \bibinfo{author}{\bibfnamefont{E.}~\bibnamefont{Innerhofer}},
  \bibinfo{author}{\bibfnamefont{H.}~\bibnamefont{M\"uller-Ebhardt}},
  \bibinfo{author}{\bibfnamefont{D.}~\bibnamefont{Ottaway}},
  \bibinfo{author}{\bibfnamefont{H.}~\bibnamefont{Rehbein}},
  \bibinfo{author}{\bibfnamefont{D.}~\bibnamefont{Sigg}},
  \bibinfo{author}{\bibfnamefont{S.}~\bibnamefont{Whitcomb}},
  \bibinfo{author}{\bibfnamefont{C.}~\bibnamefont{Wipf}}, ,
  \bibnamefont{et~al.}, \bibinfo{journal}{Phys.\ Rev.\ Lett.}
  \textbf{\bibinfo{volume}{98}}, \bibinfo{pages}{150802}
  (\bibinfo{year}{2007}).

\bibitem[{\citenamefont{Mancini et~al.}(1998)\citenamefont{Mancini, Vitali, and
  Tombesi}}]{Mancini1998}
\bibinfo{author}{\bibfnamefont{S.}~\bibnamefont{Mancini}},
  \bibinfo{author}{\bibfnamefont{D.}~\bibnamefont{Vitali}}, \bibnamefont{and}
  \bibinfo{author}{\bibfnamefont{P.}~\bibnamefont{Tombesi}},
  \bibinfo{journal}{Phys.\ Rev.\ Lett.} \textbf{\bibinfo{volume}{80}},
  \bibinfo{pages}{688} (\bibinfo{year}{1998}).

\bibitem[{\citenamefont{Cohadon et~al.}(1999)\citenamefont{Cohadon, Heidmann,
  and Pinard}}]{Cohadon1999}
\bibinfo{author}{\bibfnamefont{P.}~\bibnamefont{Cohadon}},
  \bibinfo{author}{\bibfnamefont{A.}~\bibnamefont{Heidmann}}, \bibnamefont{and}
  \bibinfo{author}{\bibfnamefont{M.}~\bibnamefont{Pinard}},
  \bibinfo{journal}{Phys. Rev. Lett.} \textbf{\bibinfo{volume}{83}},
  \bibinfo{pages}{3174} (\bibinfo{year}{1999}).

\bibitem[{\citenamefont{Kleckner and Bouwmeester}(2006)}]{Kleckner2006b}
\bibinfo{author}{\bibfnamefont{D.}~\bibnamefont{Kleckner}} \bibnamefont{and}
  \bibinfo{author}{\bibfnamefont{D.}~\bibnamefont{Bouwmeester}},
  \bibinfo{journal}{Nature} \textbf{\bibinfo{volume}{444}}, \bibinfo{pages}{75}
  (\bibinfo{year}{2006}).

\bibitem[{\citenamefont{Arcizet
  et~al.}(2006{\natexlab{b}})\citenamefont{Arcizet, Cohadon, Briant, Pinard,
  Heidmann, Mackowski, Michel, Pinard, Fran\c{c}ais, and
  Rousseau}}]{Arcizet2006a}
\bibinfo{author}{\bibfnamefont{O.}~\bibnamefont{Arcizet}},
  \bibinfo{author}{\bibfnamefont{P.-F.} \bibnamefont{Cohadon}},
  \bibinfo{author}{\bibfnamefont{T.}~\bibnamefont{Briant}},
  \bibinfo{author}{\bibfnamefont{M.}~\bibnamefont{Pinard}},
  \bibinfo{author}{\bibfnamefont{A.}~\bibnamefont{Heidmann}},
  \bibinfo{author}{\bibfnamefont{J.-M.} \bibnamefont{Mackowski}},
  \bibinfo{author}{\bibfnamefont{C.}~\bibnamefont{Michel}},
  \bibinfo{author}{\bibfnamefont{L.}~\bibnamefont{Pinard}},
  \bibinfo{author}{\bibfnamefont{O.}~\bibnamefont{Fran\c{c}ais}},
  \bibnamefont{and} \bibinfo{author}{\bibfnamefont{L.}~\bibnamefont{Rousseau}},
  \bibinfo{journal}{Phys.\ Rev.\ Lett.} \textbf{\bibinfo{volume}{97}},
  \bibinfo{pages}{133601} (\bibinfo{year}{2006}{\natexlab{b}}).

\bibitem[{\citenamefont{Poggio et~al.}(2007)\citenamefont{Poggio, Degen, Mamin,
  and Rugar}}]{Poggio2007}
\bibinfo{author}{\bibfnamefont{M.}~\bibnamefont{Poggio}},
  \bibinfo{author}{\bibfnamefont{C.~L.} \bibnamefont{Degen}},
  \bibinfo{author}{\bibfnamefont{H.~J.} \bibnamefont{Mamin}}, \bibnamefont{and}
  \bibinfo{author}{\bibfnamefont{D.}~\bibnamefont{Rugar}},
  \bibinfo{journal}{Phys.\ Rev.\ Lett.} \textbf{\bibinfo{volume}{99}},
  \bibinfo{pages}{017201} (\bibinfo{year}{2007}).

\bibitem[{\citenamefont{Courty et~al.}(2001)\citenamefont{Courty, Heidmann, and
  Pinard}}]{Courty2001}
\bibinfo{author}{\bibfnamefont{J.-M.} \bibnamefont{Courty}},
  \bibinfo{author}{\bibfnamefont{A.}~\bibnamefont{Heidmann}}, \bibnamefont{and}
  \bibinfo{author}{\bibfnamefont{M.}~\bibnamefont{Pinard}},
  \bibinfo{journal}{Eur.\ Phys.\ J.\ D} \textbf{\bibinfo{volume}{17}},
  \bibinfo{pages}{399} (\bibinfo{year}{2001}).

\bibitem[{\citenamefont{Vitali et~al.}(2002)\citenamefont{Vitali, Mancini,
  Ribichini, and Tombesi}}]{Vitali2002}
\bibinfo{author}{\bibfnamefont{D.}~\bibnamefont{Vitali}},
  \bibinfo{author}{\bibfnamefont{S.}~\bibnamefont{Mancini}},
  \bibinfo{author}{\bibfnamefont{L.}~\bibnamefont{Ribichini}},
  \bibnamefont{and} \bibinfo{author}{\bibfnamefont{P.}~\bibnamefont{Tombesi}},
  \bibinfo{journal}{Phys.\ Rev.\ A} \textbf{\bibinfo{volume}{65}},
  \bibinfo{pages}{063803} (\bibinfo{year}{2002}).

\bibitem[{\citenamefont{Tittonen et~al.}(1999)\citenamefont{Tittonen,
  Breitenbach, Kalkbrenner, M\"uller, Conradt, Schiller, Steinsland, Blanc, and
  de~Rooij}}]{Tittonen1999}
\bibinfo{author}{\bibfnamefont{I.}~\bibnamefont{Tittonen}},
  \bibinfo{author}{\bibfnamefont{G.}~\bibnamefont{Breitenbach}},
  \bibinfo{author}{\bibfnamefont{T.}~\bibnamefont{Kalkbrenner}},
  \bibinfo{author}{\bibfnamefont{T.}~\bibnamefont{M\"uller}},
  \bibinfo{author}{\bibfnamefont{R.}~\bibnamefont{Conradt}},
  \bibinfo{author}{\bibfnamefont{S.}~\bibnamefont{Schiller}},
  \bibinfo{author}{\bibfnamefont{E.}~\bibnamefont{Steinsland}},
  \bibinfo{author}{\bibfnamefont{N.}~\bibnamefont{Blanc}}, \bibnamefont{and}
  \bibinfo{author}{\bibfnamefont{N.~F.} \bibnamefont{de~Rooij}},
  \bibinfo{journal}{Phys.\ Rev.\ A} \textbf{\bibinfo{volume}{59}},
  \bibinfo{pages}{1038} (\bibinfo{year}{1999}).

\bibitem[{\citenamefont{Metzger and Karrai}(2004)}]{Metzger2004}
\bibinfo{author}{\bibfnamefont{C.~H.} \bibnamefont{Metzger}} \bibnamefont{and}
  \bibinfo{author}{\bibfnamefont{K.}~\bibnamefont{Karrai}},
  \bibinfo{journal}{Nature} \textbf{\bibinfo{volume}{432}},
  \bibinfo{pages}{1002} (\bibinfo{year}{2004}).

\bibitem[{\citenamefont{Arcizet
  et~al.}(2006{\natexlab{c}})\citenamefont{Arcizet, Briant, Heidmann, and
  Pinard}}]{Arcizet2006c}
\bibinfo{author}{\bibfnamefont{O.}~\bibnamefont{Arcizet}},
  \bibinfo{author}{\bibfnamefont{T.}~\bibnamefont{Briant}},
  \bibinfo{author}{\bibfnamefont{A.}~\bibnamefont{Heidmann}}, \bibnamefont{and}
  \bibinfo{author}{\bibfnamefont{M.}~\bibnamefont{Pinard}},
  \bibinfo{journal}{Phys.\ Rev.\ A} \textbf{\bibinfo{volume}{73}},
  \bibinfo{pages}{033819} (\bibinfo{year}{2006}{\natexlab{c}}).

\bibitem[{\citenamefont{Paternostro et~al.}(2006)\citenamefont{Paternostro,
  Gigan, Kim, Blaser, B\"ohm, and Aspelmeyer}}]{Paternostro2006}
\bibinfo{author}{\bibfnamefont{M.}~\bibnamefont{Paternostro}},
  \bibinfo{author}{\bibfnamefont{S.}~\bibnamefont{Gigan}},
  \bibinfo{author}{\bibfnamefont{M.~S.} \bibnamefont{Kim}},
  \bibinfo{author}{\bibfnamefont{F.}~\bibnamefont{Blaser}},
  \bibinfo{author}{\bibfnamefont{H.~R.} \bibnamefont{B\"ohm}},
  \bibnamefont{and}
  \bibinfo{author}{\bibfnamefont{M.}~\bibnamefont{Aspelmeyer}},
  \bibinfo{journal}{New J.\ Phys.} \textbf{\bibinfo{volume}{8}},
  \bibinfo{pages}{107} (\bibinfo{year}{2006}).

\bibitem[{\citenamefont{Vitali et~al.}(2007)\citenamefont{Vitali, Gigan,
  Ferreira, B\"ohm, Tombesi, Guerreiro, Vedral, Zeilinger, and
  Aspelmeyer}}]{Vitali2007}
\bibinfo{author}{\bibfnamefont{D.}~\bibnamefont{Vitali}},
  \bibinfo{author}{\bibfnamefont{S.}~\bibnamefont{Gigan}},
  \bibinfo{author}{\bibfnamefont{A.}~\bibnamefont{Ferreira}},
  \bibinfo{author}{\bibfnamefont{H.~R.} \bibnamefont{B\"ohm}},
  \bibinfo{author}{\bibfnamefont{P.}~\bibnamefont{Tombesi}},
  \bibinfo{author}{\bibfnamefont{A.}~\bibnamefont{Guerreiro}},
  \bibinfo{author}{\bibfnamefont{V.}~\bibnamefont{Vedral}},
  \bibinfo{author}{\bibfnamefont{A.}~\bibnamefont{Zeilinger}},
  \bibnamefont{and}
  \bibinfo{author}{\bibfnamefont{M.}~\bibnamefont{Aspelmeyer}},
  \bibinfo{journal}{Phys.\ Rev.\ Lett.} \textbf{\bibinfo{volume}{98}},
  \bibinfo{pages}{030405} (\bibinfo{year}{2007}).

\bibitem[{\citenamefont{Zhang et~al.}(2003)\citenamefont{Zhang, Peng, and
  Braunstein}}]{Zhang2003}
\bibinfo{author}{\bibfnamefont{J.}~\bibnamefont{Zhang}},
  \bibinfo{author}{\bibfnamefont{K.}~\bibnamefont{Peng}}, \bibnamefont{and}
  \bibinfo{author}{\bibfnamefont{S.~L.} \bibnamefont{Braunstein}},
  \bibinfo{journal}{Phys.\ Rev.\ A} \textbf{\bibinfo{volume}{68}},
  \bibinfo{pages}{013808} (\bibinfo{year}{2003}).

\bibitem[{\citenamefont{B\"ohm et~al.}(2006)\citenamefont{B\"ohm, Gigan,
  Blaser, Zeilinger, Aspelmeyer, Langer, B\"auerle, Hertzberg, and
  Schwab}}]{Boehm2006}
\bibinfo{author}{\bibfnamefont{H.~R.} \bibnamefont{B\"ohm}},
  \bibinfo{author}{\bibfnamefont{S.}~\bibnamefont{Gigan}},
  \bibinfo{author}{\bibfnamefont{F.}~\bibnamefont{Blaser}},
  \bibinfo{author}{\bibfnamefont{A.}~\bibnamefont{Zeilinger}},
  \bibinfo{author}{\bibfnamefont{M.}~\bibnamefont{Aspelmeyer}},
  \bibinfo{author}{\bibfnamefont{G.}~\bibnamefont{Langer}},
  \bibinfo{author}{\bibfnamefont{D.}~\bibnamefont{B\"auerle}},
  \bibinfo{author}{\bibfnamefont{J.~B.} \bibnamefont{Hertzberg}},
  \bibnamefont{and} \bibinfo{author}{\bibfnamefont{K.~C.}
  \bibnamefont{Schwab}}, \bibinfo{journal}{Appl.\ Phys.\ Lett.}
  \textbf{\bibinfo{volume}{89}}, \bibinfo{pages}{223101}
  (\bibinfo{year}{2006}).

\bibitem[{\citenamefont{Bose et~al.}(1997)\citenamefont{Bose, Jacobs, and
  Knight}}]{Bose1997}
\bibinfo{author}{\bibfnamefont{S.}~\bibnamefont{Bose}},
  \bibinfo{author}{\bibfnamefont{K.}~\bibnamefont{Jacobs}}, \bibnamefont{and}
  \bibinfo{author}{\bibfnamefont{P.~L.} \bibnamefont{Knight}},
  \bibinfo{journal}{Phys.\ Rev.\ A} \textbf{\bibinfo{volume}{56}},
  \bibinfo{pages}{4175} (\bibinfo{year}{1997}).

\bibitem[{\citenamefont{Pinard et~al.}(2005)\citenamefont{Pinard, Dantan,
  Vitali, Arcizet, Briant, and Heidmann}}]{Pinard2005b}
\bibinfo{author}{\bibfnamefont{M.}~\bibnamefont{Pinard}},
  \bibinfo{author}{\bibfnamefont{A.}~\bibnamefont{Dantan}},
  \bibinfo{author}{\bibfnamefont{D.}~\bibnamefont{Vitali}},
  \bibinfo{author}{\bibfnamefont{O.}~\bibnamefont{Arcizet}},
  \bibinfo{author}{\bibfnamefont{T.}~\bibnamefont{Briant}}, \bibnamefont{and}
  \bibinfo{author}{\bibfnamefont{A.}~\bibnamefont{Heidmann}},
  \bibinfo{journal}{Europhys. Lett.} \textbf{\bibinfo{volume}{72}},
  \bibinfo{pages}{747} (\bibinfo{year}{2005}).

\bibitem[{\citenamefont{Pirandola et~al.}(2006)\citenamefont{Pirandola, Vitali,
  Tombesi, and Lloyd}}]{Pirandola2006}
\bibinfo{author}{\bibfnamefont{S.}~\bibnamefont{Pirandola}},
  \bibinfo{author}{\bibfnamefont{D.}~\bibnamefont{Vitali}},
  \bibinfo{author}{\bibfnamefont{P.}~\bibnamefont{Tombesi}}, \bibnamefont{and}
  \bibinfo{author}{\bibfnamefont{S.}~\bibnamefont{Lloyd}},
  \bibinfo{journal}{Phys.\ Rev.\ Lett.} \textbf{\bibinfo{volume}{97}},
  \bibinfo{pages}{150403} (\bibinfo{year}{2006}).

\bibitem[{\citenamefont{Naik et~al.}(2006)\citenamefont{Naik, Buu, LaHaye,
  Armour, Clerk, Blencowe, and Schwab}}]{Naik2006}
\bibinfo{author}{\bibfnamefont{A.}~\bibnamefont{Naik}},
  \bibinfo{author}{\bibfnamefont{O.}~\bibnamefont{Buu}},
  \bibinfo{author}{\bibfnamefont{M.~D.} \bibnamefont{LaHaye}},
  \bibinfo{author}{\bibfnamefont{A.~D.} \bibnamefont{Armour}},
  \bibinfo{author}{\bibfnamefont{A.~A.} \bibnamefont{Clerk}},
  \bibinfo{author}{\bibfnamefont{M.~P.} \bibnamefont{Blencowe}},
  \bibnamefont{and} \bibinfo{author}{\bibfnamefont{K.~C.}
  \bibnamefont{Schwab}}, \bibinfo{journal}{Nature}
  \textbf{\bibinfo{volume}{443}}, \bibinfo{pages}{193} (\bibinfo{year}{2006}).

\bibitem[{\citenamefont{Thompson et~al.}(2007)\citenamefont{Thompson, Zwickl,
  Jayich, Marquardt, Girvin, and Harris}}]{Thompson2007}
\bibinfo{author}{\bibfnamefont{J.~D.} \bibnamefont{Thompson}},
  \bibinfo{author}{\bibfnamefont{B.~M.} \bibnamefont{Zwickl}},
  \bibinfo{author}{\bibfnamefont{A.~M.} \bibnamefont{Jayich}},
  \bibinfo{author}{\bibfnamefont{F.}~\bibnamefont{Marquardt}},
  \bibinfo{author}{\bibfnamefont{S.~M.} \bibnamefont{Girvin}},
  \bibnamefont{and} \bibinfo{author}{\bibfnamefont{J.~G.~E.}
  \bibnamefont{Harris}}, \bibinfo{journal}{arXiv:0707.1724}
  (\bibinfo{year}{2007}).

\end{thebibliography}
\end{document}